\documentclass[12pt]{JHEP3}
\usepackage{graphics,latexsym,amsmath,epsfig,bm,cite}

\newcommand{\as}{\alpha_s}

\renewcommand{\d}{\mathrm{d}}
\newcommand{\A}{\mathcal{A}}

\newcommand{\C}{\mathbf{C}}

\newcommand{\ldot}{\!\cdot\!}
\newcommand{\sgn}{\mathrm{sgn}}
\newcommand{\smfrac}[2]{{\textstyle\frac{#1}{#2}}}
\renewcommand{\vec}[1]{\underline{#1}}
\renewcommand{\S}{\mathrm{S}}

\newcommand{\R}{\mathrm{R}}

\newcommand{\lev}[1]{\lambda^{(#1)}}

\newcommand{\comt}[1]{}

\title{The colour evolution of the process~\boldmath$q q \to q q g$}
\author{A. Kyrieleis \\School of Physics \& Astronomy, University of Manchester, \\
  Oxford Road, Manchester M13 9PL, U.K.\\
 \email{kyrieleis@hep.man.ac.uk}}
\author{M.H. Seymour \\School of Physics \& Astronomy, University of Manchester
\\and \\
Theoretical Physics Group (PH-TH), CERN, CH-1211 Geneva 23,
Switzerland\\
\email{Mike.Seymour@cern.ch}}

\abstract{We calculate the soft anomalous dimension matrix for a
  five-parton process, $qq\to~qqg$. Considering different bases we
  unveil some interesting properties of this matrix.}

\keywords{qcd, jet}
\preprint{CERN-PH-TH/2005-188}

\begin{document}
\noindent
\section{Introduction}
The understanding of 
gaps-between-jets processes has been
 subject to great progress over  the last few years. Of central
 importance in this context  is the energy flow into the 
interjet region as a very useful observable in the description of
 gaps-between-jets processes.  
According to the method of Sterman et al.\
\cite{Collins88,Sotir93,Laenen96, Kidon98}, the cross section for
interjet energy flow can be `refactorized' into hard and soft parts at
some factorization scale $\mu$.  For all but the most trivial processes,
these parts have a matrix structure in the space of colour flows of the
hard process, with the hard amplitude represented by a vector $\vec\A$ in
this space and the sum of all possible soft corrections to the cross
section represented by a matrix $\S$.  The $\mu$ dependence of each is
accounted for by an anomalous dimension $\Gamma$, also a matrix in
colour space.  Logarithms of the ratio of the hard and soft scales can
be summed to all orders by taking $\mu$ of order the hard scale in the
hard amplitude and the soft scale in the soft matrix, using the
exponential of the integral of $\Gamma$ to connect the two
scales. The anomalous dimension matrix $\Gamma$ has been calculated
for all (QCD) $2\to 2$ processes and  for various definitions of the
 final  state \cite{OdSt98,  Od00, BeKuSt, AbSey03}.  

It is the purpose of this paper to calculate for the first time the
anomalous dimension matrix for a $2\to3$ process, namely $qq\to qqg$;
to distinguish it from the one for the $2\to2$ process $qq\to qq$, we
denote it $\Lambda$.
 This can serve as a starting
point to  improve the  
understanding of theoretical aspects in the description of gaps-between-jets
processes. The calculation of this matrix is also a first step towards
energy flow analyses
of 3-jet processes which are particularly interesting  at the LHC. 

In this paper we  consider the process  $qq\to qqg$ with a gap defined by a central
rapidity region of length  $Y<\Delta y$ where $\Delta y$ is the
rapidity separation of the outgoing quarks. The real gluon is restricted
to the region outside the gap.
$\Lambda$~is then obtained by calculating the virtual
corrections to this process from a  
softer gluon connecting the external lines in all possible
ways.  In common with other calculations of gaps-between-jets cross
sections \cite{OdSt98,  Od00, BeKuSt, AbSey03}, we assume a perfect
real--virtual cancellation outside the gap region.  Thus the virtual
gluon is integrated only over the rapidity interval of the gap and over
all azimuthal angles.

We represent the result for $\Lambda$  in three colour
bases and thereby  shed light on different aspects of it.

\section{The anomalous dimension matrix for \boldmath$qq\to qqg$}
Let us start with the colour structure. 
For the $q_iq_j\to q_kq_lg_a$ system (where the subscripts are the
colour indices) there are four independent colour states
needed. We first choose the $t$-channel basis
\begin{eqnarray}
  \C_1 &=& T^a_{ki}\delta_{lj} + \delta_{ki}T^a_{lj},\label{eq:b1} \\
  \C_2 &=& T^b_{ki}T^c_{lj}\,d^{abc}, \\
  \C_3 &=& T^a_{ki}\delta_{lj} - \delta_{ki}T^a_{lj}, \\
  \C_4 &=& T^b_{ki}T^c_{lj}\,i\!f^{abc}\label{eq:b4}.
\end{eqnarray}
The lowest order soft matrix (which contains the traces of the  squared
operators of
the basis) is given in this basis by
\begin{equation}
  \S = \left(\begin{array}{cccc}
    N_c(N_c^2-1) & 0 & 0 & 0 \\
    0 & \frac1{4N_c}(N_c^2-1)(N_c^2-4) & 0 & 0 \\
    0 & 0 & N_c(N_c^2-1) & 0 \\
    0 & 0 & 0 & \frac14N_c(N_c^2-1)
  \end{array}\right).
\end{equation}
The momenta of the hard process are labeled in the following way
\begin{align}
q(p_1) + q(p_2) \to q(p_3) + q(p_4) + g(k).
\label{eq:hard}
\end{align} 
We work in the frame in which 1 and 2 collide head on and the gap region
is central in rapidity,
\begin{eqnarray}
  p_1 &=& E_1(1;0,0,\phantom{-}1), \\
  p_2 &=& E_2(1;0,0,-1), \\
  p_3 &=& q_{\perp3}\left(\cosh y_3;
    \phantom{\sin{}}\,0,\phantom{\cos{}}\,1,\sinh y_3\right), \\
  p_4 &=& q_{\perp4}\left(\cosh y_4;
    \sin\varphi,\cos\varphi,\sinh y_4\right), \\
  k &=& k_{\perp\phantom{4}}\left(\cosh y_{\phantom{4}};
    \sin\phi,\cos\phi,\sinh y_{\phantom{4}}\right).
\end{eqnarray}
Note that in the limit in which the emitted gluon is much softer than
the quarks, $k_\perp\ll q_{\perp3,4}$, momentum conservation implies
$q_{\perp3}=q_{\perp4}$ and $\varphi=\pi$ and the kinematics are
identical to the lowest order process $qq\to qq$.  We are interested in
the case that the quark jets are either side of the gap and can
therefore assume $y_3>0$ and $y_4<0$.

We denote the rapidity, azimuthal angle and  transverse momentum
of the virtual gluon $k'$ by $y', \phi'$ and $k'_\perp$, respectively.
For future use, we define 
\begin{equation}
  s_y = \sgn(y).
\end{equation}
The gap is defined by a central rapidity region of length $Y$.  Since we
are interested in the case of $k$ outside the gap and $k'$ within it, we
have
\begin{equation}
  |y'| < \frac Y2 < |y|
\end{equation}
and hence
\begin{equation}
  \sgn(y-y')=\sgn(y)=s_y.
\end{equation}
We denote the hard amplitude \eqref{eq:hard} evaluated at
refactorization scale $\mu$ by the (four dimensional) vector
$\vec\A(\mu)$.  The anomalous dimension matrix $\Lambda$ is then defined
through the evolution of $\vec\A(\mu)$,
\begin{equation}
  \label{eq:LambdaDef}
  \mu\frac{\d}{\d\mu}\vec\A = \frac{2\as}{\pi} \;\Lambda\;\vec\A.
\end{equation}
We can extract $\Lambda$ from a one-loop calculation by expanding
\eqref{eq:LambdaDef} to leading order,
\begin{align}
\vec\A^{(1)}=-\frac{2\as}{\pi}\int \frac{dk'_\perp}{k'_\perp} \;\Lambda
\;\vec\A^{(0)},
\label{eq:def}
\end{align}
where $\vec\A^{(0)}$ and $\vec\A^{(1)}$ are respectively the lowest
order and one-loop amplitudes.  The latter is calculated from the
virtual corrections to the hard process from a gluon coupling
two external lines in all possible ways.  We work in the eikonal
effective theory. The region of integration is 
\begin{align}
  0<&\;\phi'<2 \pi\\
  -Y/2 < &\;y'<Y/2.
\end{align}
Details of the calculation of $\Lambda$ can be found in the
Appendix. We only state the result here written as a sum of four parts.

\begin{eqnarray}
  \hspace*{-2em}
  \Lambda &=& \left(\begin{array}{cccc}
    \frac{N_c}4(Y-i\pi)+\frac1{2N_c}i\pi &
    (\frac14-\frac1{N_c^2})i\pi &
    -\frac{N_c}4s_yY &
    0
    \\
    i\pi &
    \frac{N_c}4(2Y-i\pi)-\frac3{2N_c}i\pi &
    0 &
    0
    \\
    -\frac{N_c}4s_yY &
    0 &
    \frac{N_c}4(Y-i\pi)-\frac1{2N_c}i\pi &
    -\frac14i\pi
    \\
    0 &
    0 &
    -i\pi &
    \frac{N_c}4(2Y-i\pi)-\frac1{2N_c}i\pi
  \end{array}\right)
  \hspace*{-2em}\nonumber\\\hspace*{-2em}&&+
  \left(\begin{array}{cccc}
  N_c & 0 & 0 & 0 \\ 0 &
  N_c & 0 & 0 \\ 0 & 0 &
  N_c & 0 \\ 0 & 0 & 0 &
  N_c
  \end{array}\right)\frac14\rho(2|y|)
  \hspace*{-2em}\nonumber\\\hspace*{-2em}&&+
  \left(\begin{array}{cccc}
  C_F & 0 & 0 & 0 \\ 0 &
  C_F & 0 & 0 \\ 0 & 0 &
  C_F & 0 \\ 0 & 0 & 0 &
  C_F
  \end{array}\right)\left(\frac14\rho(2|y_3|)+\frac14\rho(2|y_4|)\right)
  \hspace*{-2em}\nonumber\\\hspace*{-2em}&&+
  \left(\begin{array}{cccc}
    \frac{N_c}4(-\frac12\lambda) &
    0 &
    \frac{N_c}4(-\frac12s_y\lambda) &
    \frac14(\frac12s_y\lambda)
    \\
    0 &
    \frac{N_c}4(-\frac12\lambda) &
    0 &
    \frac{N_c}4(\frac12s_y\lambda)
    \\
    \frac{N_c}4(-\frac12s_y\lambda) &
    0 &
    \frac{N_c}4(-\frac12\lambda) &
    \frac14(-\frac12\lambda)
    \\
    \frac12s_y\lambda &
    (\frac{N_c}4-\frac1{N_c})(\frac12s_y\lambda) &
    -\frac12\lambda &
    \frac{N_c}4(-\frac12\lambda)
  \end{array}\right),\label{eq:lambda}
\end{eqnarray}
where we have defined
\begin{align}
  \rho(y)&\equiv
  \log\frac{\sinh(y/2+Y/2)}{\sinh(y/2-Y/2)}-Y,\\
\lambda &\equiv
   \frac12\log
  \frac{\cosh(|\bar y| + |y| +Y)-\cos({\bar \phi})}{\cosh(|\bar y|
  + |y|-Y)-\cos({\bar \phi})}-Y
\end{align}
with  
\begin{equation}
  \bar\phi \equiv \left\{ \begin{array}{rl}
    \phi         &\; y>0, \\
    \varphi-\phi &\; y<0,
  \end{array} \right.
\end{equation}
and
\begin{equation}
  \bar y \equiv \left\{ \begin{array}{rl}
    y_3 &\; y>0, \\
    y_4 &\; y<0.
  \end{array} \right.
\end{equation}
We have grouped the four terms of $\Lambda$ in the following way.  For
fixed $Y$, the first line contains all the terms that remain when
$y,y_3,y_4\to\pm\infty$ (the high energy limit), the second line
contains the additional terms that remain when $y$ is finite, the third line
contains the additional terms that remain when $y_3$ or $y_4$ is finite 
and the last line contains the additional terms that remain when
$y_3$, $y_4$ and $y$ are finite.
For future reference we
define these four lines  to be
$\Lambda_{1,2,3,4}$ respectively.

 For reasons that will become apparent
shortly, it will be useful to modify the matrix $\Lambda$. 
 Adding a multiple of the
identity matrix to any matrix does not change its eigenvectors and
simply adds a constant to all of its eigenvalues.  Moreover, adding an
imaginary constant to all the eigenvalues of $\Lambda$ will not change
the physics, since the energy dependence comes from combinations ${\lev
i}^*+\lev j$.  Therefore we are free to add any imaginary constant times
the identity matrix to $\Lambda$.  From now on we shall denote
$\Lambda$ the matrix  obtained from 
\eqref{eq:lambda} by
\begin{equation}
\Lambda  \to \Lambda + N_c/4\,i\pi.\label{eq:shift}
\end{equation}
The eigenvalues of this matrix $\Lambda$   are: 
\begin{align}
  \lev1 =& \frac{N_c}2Y+\frac{N_c-1}{2N_c}i\pi
  +N_c\smfrac14\rho(2|y|)
  +C_F\left(\smfrac14\rho(2|y_3|)+\smfrac14\rho(2|y_4|)\right)
  +\smfrac14\lambda, \label{eq:ev1}\\
  \lev2 =& \frac{N_c}2Y-\frac{N_c+1}{2N_c}i\pi
  +N_c\smfrac14\rho(2|y|)
  +C_F\left(\smfrac14\rho(2|y_3|)+\smfrac14\rho(2|y_4|)\right)
  -\smfrac14\lambda, \\
  \lev3 =&
  \frac{N_c^2Y-2 i\pi-N_c\sqrt{N_c^2Y^2-4Yi\pi-4\pi^2}}{4N_c}\nonumber\\
  &\hspace*{3cm} +N_c\smfrac14\rho(2|y|)
  +C_F\left(\smfrac14\rho(2|y_3|)+\smfrac14\rho(2|y_4|)\right)
  -\smfrac{N_c}4\lambda, \\
  \lev4 =& \frac{N_c^2Y-2i\pi+N_c\sqrt{N_c^2Y^2-4Yi\pi-4\pi^2}}{4N_c}\nonumber\\
  &\hspace*{3cm} +N_c\smfrac14\rho(2|y|)
  +C_F\left(\smfrac14\rho(2|y_3|)+\smfrac14\rho(2|y_4|)\right)
  -\smfrac{N_c}4\lambda\label{eq:ev4}
\end{align}
and it is diagonalized by 
\begin{equation}
  \R = \left(\begin{array}{cccc}
    \frac12s_y& -\frac12s_y& (-\frac3{2N_c}-i\frac{N_cY}{2\pi}
    +\frac{i\lev3}\pi)s_y&
    (-\frac1{2N_c}-\frac{i\lev3}\pi)s_y\\
    \frac{N_c}{N_c+2}s_y& \frac{N_c}{N_c-2}s_y& -s_y& -s_y\\
    -\frac12& \frac12& -\frac1{2N_c}-i\frac{N_cY}{2\pi}+\frac{i\lev3}\pi&
    \frac1{2N_c}-\frac{i\lev3}\pi\\
    1& 1& 1& 1
  \end{array}\right).
\end{equation}
Note that whereas $\Lambda_1$ has four different eigenvalues, two
eigenvalues of $\Lambda_4$ are degenerate.
\subsection[Block Diagonalization of $\Lambda$]
           {Block Diagonalization of \boldmath$\Lambda$}
The anomalous dimension matrix $\Gamma$ for the hard process $qq\to qq$
is defined in  exact analogy to $\Lambda$, \eqref{eq:def}. In the high
energy limit ($|y_{3,4}|\to\infty$) and in the $t$-channel singlet--octet
basis it reads:
\begin{equation}
  \Gamma = \left(\begin{array}{cc}
    0\: & \left(\frac14-\frac1{4N_c^2}\right)i\pi \\ i\pi\: &
    \frac{N_c}2Y-\frac1{N_c}(i\pi)
  \end{array}
  \right).
\end{equation}
Two of the eigenvalues of $\Lambda_1$ coincide with the
eigenvalues of $\Gamma$ (to enable this was the reason for the
modification \eqref{eq:shift} of $\Lambda$). We can therefore construct a matrix 
\begin{equation}
  \hat{\R} = \sqrt{\frac{N_c}{2(N_c^2-1)}}
  \left(\begin{array}{cccc}
    \frac12s_y&-\frac12s_y&s_y&\frac1{2N_c}s_y\\
    \frac{N_c}{N_c+2}s_y&\frac{N_c}{N_c-2}s_y&0&s_y\\
    -\frac12&\frac12&1&-\frac1{2N_c}\\
    1&1&0&-1
  \end{array}\right).
\end{equation}
which diagonalizes $\Lambda_4$ and transforms $\Lambda_1$ in the
following way:
\begin{equation}
  \hat{\R}^{-1}\Lambda_1\hat{\R}=\left(\begin{array}{cccc}
    \lev1_1&0&0&0\\
    0&\lev2_1&0&0\\
    0&0&0&\left(\frac14-\frac1{4N_c^2}\right)i\pi\\
    0&0&i\pi&\frac{N_c}2Y-\frac1{N_c}(i\pi)
  \end{array}\right).
\end{equation}
where $\lambda_1^{(i)}$ are the eigenvalues of $\Lambda_1$, which can
be obtained from (\ref{eq:ev1}-\ref{eq:ev4}) by setting $\rho$ and $\lambda$ to 0.
Not only is this matrix block diagonal but, remarkably, the upper left block is
itself diagonal, and the lower right block is identical to $\Gamma$.

Note that the soft matrix in this basis,  
\begin{eqnarray}
  \hat{\R}^\dagger\S\hat{\R}
  &=& \left(\begin{array}{cccc}
    \frac{N_c^2}2\,\frac{N_c+1}{N_c+2}&0&0&0\\
    0&\frac{N_c^2}2\,\frac{N_c-1}{N_c-2}&0&0\\
    0&0&N_c^2&0\\
    0&0&0&\frac14(N_c^2-1)
  \end{array}\right),
\end{eqnarray}
is still diagonal and that its
lower right block is identical to the soft matrix of the $2\to2$ process
(the latter property is the reason for our choice of normalization
for~$\hat{\R}$).  It is also interesting to note that in this basis the
anomalous dimension matrix $\Lambda$ is $s_y$ independent and that all
of the $s_y$ dependence is carried by the definitions of the basis
states, which are different for $s_y=\pm1$.

\subsection[The $s$-channel basis]{The \boldmath$s$-channel basis}
Reference \cite{Dokshitzer:2005ek} advocated using the set of
$s$-channel projectors as the colour basis for $2\to2$ processes.  In
this section we present our results for $\Lambda$ in an alternative
block-diagonal form in which its lower right block is identical to
$\Gamma$ transformed into the $s$-channel basis and show that its basis
states have a simple form.

\renewcommand{\P}{\mathbf{P}}
For a $qq$ state, the projectors are
\begin{eqnarray}
  \P_{\bar 3} &=& \frac12
  \Bigl(\delta_{ki}\delta_{lj}-\delta_{li}\delta_{kj}\Bigr), \\
  \P_6        &=& \frac12
  \Bigl(\delta_{ki}\delta_{lj}+\delta_{li}\delta_{kj}\Bigr).
\end{eqnarray}
We can transform between the $t$-channel basis we have been using so far
and the $s$-channel basis using the matrix
\begin{equation}
  \R_{st} = \left(\begin{array}{cc}
    \frac{N_c-1}{2N_c}&\frac{N_c+1}{2N_c}\\
    -1&+1
  \end{array}\right).
\end{equation}
That is, $\Gamma$ transforms to
\begin{equation}
  \R_{st}^{-1}\,\Gamma\,\R_{st} =
  \left(\begin{array}{cc}
    \frac14(N_c+1)Y-\frac{N_c+1}{2N_c}i\pi&
    -\frac14(N_c+1)Y\\
    -\frac14(N_c-1)Y&
    \frac14(N_c-1)Y+\frac{N_c-1}{2N_c}i\pi
  \end{array}\right)
  \label{eq:schannelGamma}
\end{equation}
in the high energy limit and $\S$ to
\begin{equation}
  \R_{st}^\dagger\,\S\,\R_{st} =
  \left(\begin{array}{cc}
    \frac12N_c(N_c-1)&0\\0&\frac12N_c(N_c+1)
  \end{array}\right).
\end{equation}
Note that the imaginary terms appear only in the diagonal of $\Gamma$
and that the entries in $\S$ correspond to the multiplicities of the
basis states, 3 and 6 for $N_c=3$, two of the advantages of the
$s$-channel projector basis.

We wish to express $\Lambda$ in a block diagonal form in which the
bottom right block is equal to $\Gamma$ in the $s$-channel basis,
Eq.~(\ref{eq:schannelGamma}).  To this end we define a matrix
\begin{equation}
  \hat\R_{st} = \left(\begin{array}{cccc}
    \sqrt{\frac{N_c+2}{N_c}}&0&0&0\\
    0&\sqrt{\frac{N_c-2}{N_c}}&0&0\\
    0&0&\frac{N_c-1}{2N_c}&\frac{N_c+1}{2N_c}\\
    0&0&-1&+1
  \end{array}\right),
\end{equation}
in which the bottom right block is equal to $\R_{st}$ and the diagonal
entries in the top left block are arbitrary~-- the particular choice
made here will lead to a convenient result for the soft matrix.

Transforming $\Lambda_1$ and $\S$ from the original $t$-channel basis,
we obtain
\begin{equation}
  \label{eq:Lambdafinal}
  \hat\R_{st}^{-1}\,\hat\R^{-1}\,\Lambda_1\,\hat\R\,\hat\R_{st}
  =\left(\begin{array}{cccc}
    \lev1_1&0&0&0\\
    0&\lev2_1&0&0\\
    0&0&
    \frac14(N_c+1)Y-\frac{N_c+1}{2N_c}i\pi&
    -\frac14(N_c+1)Y\\
    0&0&
    -\frac14(N_c-1)Y&
    \frac14(N_c-1)Y+\frac{N_c-1}{2N_c}i\pi
  \end{array}\right)
\end{equation}
and
\begin{equation}
  \label{eq:Sfinal}
  \hat\R_{st}^\dagger\,\hat\R^\dagger\,\S\,\hat\R\,\hat\R_{st}
  =\left(\begin{array}{cccc}
    \frac12N_c(N_c+1)&0&0&0\\
    0&\frac12N_c(N_c-1)&0&0\\
    0&0&\frac12N_c(N_c-1)&0\\
    0&0&0&\frac12N_c(N_c+1)
  \end{array}\right).
\end{equation}
By construction, the lower right block of
$\Lambda$ is equal to $\Gamma$ in the high energy limit, the lower right
block of $\S$ is equal to $\S$ in the $2\to2$ $s$-channel basis and the
upper left  entries
of $\Lambda$ are left unchanged.  The upper left entries of $\S$ are set
by our arbitrary choices in $\hat\R_{st}$ for reasons that will be seen
shortly.
Beyond the high energy limit, the bottom right block of $\Lambda$
contains a term $\frac14C_F(\rho(2|y_3|)+\rho(2|y_4|)) +
\frac14N_c(\rho(2|y|)-\lambda)$ times the identity matrix while $\Gamma$
contains just a term $\frac14C_F(\rho(2|y_3|)+\rho(2|y_4|))$ times the identity
matrix, resulting in a small mismatch.

The actual definitions of the basis states can be read off from the
columns of $\hat\R\,\hat\R_{st}$ and can be written in forms
proportional to the $s$-channel projectors for the incoming quarks.
Since the matrix $\hat\R$ depends on $s_y$, these states are different for
$s_y=\pm1$ (recall that $s_y=+1$ implies that the gluon is on the same side
of the rapidity gap as incoming quark $i$ and outgoing quark $k$, while
$s_y=-1$ implies that it is on the other side).  For $s_y=+1$ we have
\begin{eqnarray}
  \hat\C_1 &\propto&
  \frac12\Bigl(\delta_{mi}\delta_{nj}+\delta_{ni}\delta_{mj}\Bigr)
  \Bigl(\delta_{km}T_{ln}^a-\frac1{N_c+1}T_{km}^a\delta_{ln}
    \Bigr), \\
  \hat\C_2 &\propto&
  \frac12\Bigl(\delta_{mi}\delta_{nj}-\delta_{ni}\delta_{mj}\Bigr)
  \Bigl(\delta_{km}T_{ln}^a+\frac1{N_c-1}T_{km}^a\delta_{ln}
    \Bigr), \\
  \hat\C_3 &\propto&
  \frac12\Bigl(\delta_{mi}\delta_{nj}-\delta_{ni}\delta_{mj}\Bigr)
  \Bigl(T_{km}^a\delta_{ln}\Bigr), \\
  \hat\C_4 &\propto&
  \frac12\Bigl(\delta_{mi}\delta_{nj}+\delta_{ni}\delta_{mj}\Bigr)
  \Bigl(T_{km}^a\delta_{ln}\Bigr),
\end{eqnarray}
while for $s_y=-1$ we have
\begin{eqnarray}
  \hat\C_1 &\propto&
  \frac12\Bigl(\delta_{mi}\delta_{nj}+\delta_{ni}\delta_{mj}\Bigr)
  \Bigl(T_{km}^a\delta_{ln}-\frac1{N_c+1}\delta_{km}T_{ln}^a
    \Bigr), \\
  \hat\C_2 &\propto&
  \frac12\Bigl(\delta_{mi}\delta_{nj}-\delta_{ni}\delta_{mj}\Bigr)
  \Bigl(T_{km}^a\delta_{ln}+\frac1{N_c-1}\delta_{km}T_{ln}^a
    \Bigr), \\
  \hat\C_3 &\propto&
  \frac12\Bigl(\delta_{mi}\delta_{nj}-\delta_{ni}\delta_{mj}\Bigr)
  \Bigl(\delta_{km}T_{ln}^a\Bigr), \\
  \hat\C_4 &\propto&
  \frac12\Bigl(\delta_{mi}\delta_{nj}+\delta_{ni}\delta_{mj}\Bigr)
  \Bigl(\delta_{km}T_{ln}^a\Bigr).
\end{eqnarray}

It is important to note that although we have made arbitrary choices
that affect the normalizations of these states in order to get $\S$ into
the form Eq.~(\ref{eq:Sfinal}), their forms are otherwise determined
entirely by the physics of $\Lambda$.  We see that the two states that
evolve  like a $qq\to qq$ system have a colour structure given by
the $qq\to qq$ projectors followed by a gluon emission from the outgoing
quark it is on the same side of the gap as.  The two
other states are
given similarly by projectors followed by an emission from the other
outgoing quark, up to colour-suppressed terms coming from emission on
the same side.

\section{Conclusions}
We have calculated the anomalous dimension matrix $\Lambda$ for the
five-parton process $qq\to qqg$ and presented it in several different
colour bases.  It seems likely that the generalization of the
$s$-channel basis, Eq.~(\ref{eq:Lambdafinal}), will be most useful both for
gaining insight into the physics of $\Lambda$ and for performing
practical calculations.  We anticipate using $\Lambda$ to improve the
theoretical understanding of gaps-between-jets processes and ultimately
to calculate energy flow observables in 3-jet processes, which are
particularly interesting at the LHC.  The latter however requires the
anomalous dimension matrices for all 3-jet processes to be calculated, a
highly non-trivial problem: in the most complicated case of $gg\to ggg$
one expects up to 44 independent colour amplitudes and a deeper
theoretical insight seems necessary to organize the calculation.  In
particular it would be extremely interesting to see whether the block
diagonal structure we found for $qq\to qqg$ can be generalized to
arbitrary processes, with one block always equal to the anomalous
dimension matrix of a lower-order process.

\appendix
\section{Calculation of \boldmath$\Lambda$}
In the basis (\ref{eq:b1}-\ref{eq:b4}) $\Lambda$ has the following
colour structure
\begin{eqnarray}
  \hspace*{-2em}
  \Lambda &=& \scalebox{0.7}{$\displaystyle\left(\begin{array}{cccc}
    \frac1{2N_c}(\Omega_{12}+\Omega_{34}+\Omega_{14}+\Omega_{23}) &
    (\frac14-\frac1{N_c^2})
      (\Omega_{12}+\Omega_{34}+\Omega_{14}+\Omega_{23}) &
    0 &
    \frac14(-\Omega_{14}+\Omega_{23})
    \\
    &&&\\
    \Omega_{12}+\Omega_{34}+\Omega_{14}+\Omega_{23} &
    -\frac3{2N_c}(\Omega_{12}+\Omega_{34}+\Omega_{14}+\Omega_{23})
     &
    0 &
    \frac{N_c}4(-\Omega_{14}+\Omega_{23})
    \\
    &+\frac{N_c}4(\Omega_{14}+\Omega_{23})&&\\
    0 &
    0 &
    -\frac1{2N_c}(\Omega_{12}+\Omega_{34}+\Omega_{14}+\Omega_{23}) &
    \frac14(-\Omega_{12}+\Omega_{34})
    \\
    &&&\\
    -\Omega_{14}+\Omega_{23} &
    (\frac{N_c}4-\frac1{N_c})(-\Omega_{14}+\Omega_{23}) &
    -\Omega_{12}+\Omega_{34} &
    -\frac1{2N_c}(\Omega_{12}+\Omega_{34}+\Omega_{14}+\Omega_{23})
    \\
    &&& +\frac{N_c}4(\Omega_{14}+\Omega_{23})
  \end{array}\right)$}
  \hspace*{-2em}\nonumber\\[.3cm]\hspace*{-2em}&&+
  \scalebox{0.7}{$\displaystyle\left(\begin{array}{cccc}
    \frac{N_c}4(\Omega_{1k}-\Omega_{2k}-\Omega_{3k}+\Omega_{4k}) &
    0 &
    \frac{N_c}4(\Omega_{1k}+\Omega_{2k}-\Omega_{3k}-\Omega_{4k}) &
    \frac14(\Omega_{1k}+\Omega_{2k}+\Omega_{3k}+\Omega_{4k})
    \\
    0 &
    \frac{N_c}4(\Omega_{1k}-\Omega_{2k}-\Omega_{3k}+\Omega_{4k}) &
    0 &
    \frac{N_c}4(\Omega_{1k}+\Omega_{2k}+\Omega_{3k}+\Omega_{4k})
    \\
    \frac{N_c}4(\Omega_{1k}+\Omega_{2k}-\Omega_{3k}-\Omega_{4k}) &
    0 &
    \frac{N_c}4(\Omega_{1k}-\Omega_{2k}-\Omega_{3k}+\Omega_{4k}) &
    \frac14(-\Omega_{1k}+\Omega_{2k}-\Omega_{3k}+\Omega_{4k})
    \\
    \Omega_{1k}+\Omega_{2k}+\Omega_{3k}+\Omega_{4k} &
    (\frac{N_c}4-\frac1{N_c})(\Omega_{1k}+\Omega_{2k}+\Omega_{3k}+\Omega_{4k}) &
    -\Omega_{1k}+\Omega_{2k}-\Omega_{3k}+\Omega_{4k} &
    \frac{N_c}4(\Omega_{1k}-\Omega_{2k}-\Omega_{3k}+\Omega_{4k})
  \end{array}\right)$}
  \hspace*{-2em}\nonumber\\[.3cm]\hspace*{-2em}&&+
\scalebox{0.7}{$\displaystyle\left(\begin{array}{cccc}
    (\frac{N_c}4-\frac1{2N_c})(\Omega_{13}+\Omega_{24}) & 0 &
    \frac{N_c}4(-\Omega_{13}+\Omega_{24}) & 0 \\
    0 & -\frac1{2N_c}(\Omega_{13}+\Omega_{24}) & 0 & 0 \\
    \frac{N_c}4(-\Omega_{13}+\Omega_{24}) & 0 &
    (\frac{N_c}4-\frac1{2N_c})(\Omega_{13}+\Omega_{24}) & 0 \\
    0 & 0 & 0 & -\frac1{2N_c}(\Omega_{13}+\Omega_{24})
  \end{array}\right)$}.\label{eq:colstruc}
\end{eqnarray}
$\Omega_{ij}$ corresponds to the case in which the 
virtual gluon couples quarks $i$ and $j$. $\Omega_{ik}$ accounts for the
coupling of quark $i$ and the gluon $k$. These functions are given by
\begin{equation}
  \Omega_{ij} = \frac12\delta_i\delta_j
  \left[\int_\Omega \d y' \frac{\d\phi'}{2\pi}\;\omega_{ij}
    -\smfrac12(1-\delta_i\delta_j)i\pi\right]
  \label{Omegadef}
\end{equation}
and 
\begin{equation}
  \Omega_{ik} = \frac12\delta_i\delta_k\delta_g
  \left[\int_\Omega \d y' \frac{\d\phi'}{2\pi}\;\omega'_{ik}
    -\smfrac12(1-\delta_i\delta_k)i\pi\right]\label{eq:Oij}
\end{equation}
where we have introduced the shorthands
\begin{equation}
  \omega_{ij} = \frac{\frac12k'^2_\perp\,p_i\ldot p_j}{p_i\ldot
  k'\,k'\ldot p_j}\;, \qquad   \omega'_{ik} = \frac{\frac12k'^2_\perp\,p_i\ldot k}{p_i\ldot k'\,k'\ldot k}.\label{eq:Oik}
\end{equation}
We have $\delta_i\delta_j=-1$ if $i$ and $j$ are both incoming or both
outgoing and $+1$ otherwise.  $\delta_g$~depends on the topology of the
triple-gluon vertex: in our convention in which the indices of
$if^{abc}$ are labeled in an anticlockwise direction around the vertex,
if, with the vertex rotated so that the momentum of the eikonal gluon is
flowing horizontally from left to right, the soft gluon is above it,
then $\delta_g=+1$, and if below, $\delta_g=-1$.

It is worth pointing out that in the general case, the elements of
$\Lambda$, Eq.~\eqref{eq:colstruc} obey the following equality:
\begin{equation}
  \Lambda_{ij}/\S_{jj}=\Lambda_{ji}/\S_{ii} \quad\mbox{(no sum)},
\end{equation}
implying that in an orthonormal basis in which $\S$ is equal to the
identity matrix, $\Lambda$ is a symmetric matrix.  This property is
therefore valid independently of the observable to which $\Lambda$
contributes.  It was
pointed out in Ref.~\cite{Seymour:2005ze} that this property is true of
all anomalous dimension matrices that have been calculated to date,
although no explanation of this fact was offered.

Carrying out
the integrations in (\ref{eq:Oij}, \ref{eq:Oik}) over the region
$\Omega$:
\begin{align}
  0<&\;\phi'<2 \pi\\
  -Y/2 < &\;y'<Y/2,
\end{align}
we obtain
\begin{eqnarray}
  \Omega_{12} &=& -\frac12(Y-i\pi), \\
  \Omega_{34} &=& -\frac12(Y-i\pi)-\frac14\rho(2|y_3|)-\frac14\rho(2|y_4|), \\
  \Omega_{14} &=& \frac12Y+\frac14\rho(2|y_4|), \\
  \Omega_{23} &=& \frac12Y+\frac14\rho(2|y_3|), \\
  \Omega_{13} &=& \frac14\rho(2|y_3|),\\
  \Omega_{24} &=& \frac14\rho(2|y_4|),\\
  \Omega_{1k} &=& \phantom{-}\frac14(1-s_y)Y+\frac14\rho(2|y|), \\
  \Omega_{2k} &=& -\frac14(1+s_y)Y-\frac14\rho(2|y|),\\
  \Omega_{3k} &=& -\frac12
  \left[
    \frac12(1-s_y)Y
   +\frac12\rho(2|y_3|)
   +\frac12\rho(2|y|) \right.\nonumber\\
  &&\left.\qquad\quad  -\frac12(1+s_y)\lambda(|y_3|+|y|, \phi)
    -i\pi\right], \\
  \Omega_{4k} &=& \phantom{-}\frac12
  \left[
   \frac12(1+s_y)Y
   +\frac12\rho(2|y_4|)
   +\frac12\rho(2|y|)\right.\nonumber\\
  &&\left.\qquad\quad -\frac12(1-s_y)\lambda(|y_4|+|y|, \varphi-\phi)
    -i\pi\right].
\end{eqnarray}
where we have defined
\begin{align}
  s_y &= \sgn(y),\\
  \rho(y)&\equiv
  \log\frac{\sinh(y/2+Y/2)}{\sinh(y/2-Y/2)}-Y,\\
\lambda(y,\phi) &\equiv
   \frac12\log
  \frac{\cosh(y +Y)-\cos(\phi)}{\cosh(y-Y)-\cos(\phi)}-Y.
\end{align}
It is useful to note that 
\begin{equation}
  \lambda(y,0) = \rho(y).
\end{equation}

\acknowledgments
We are extremely grateful to Jeff Forshaw for interesting discussions of
this and related topics.  AK is supported by PPARC grant PP/D000157/1.


\end{document}